\documentstyle[12pt]{article}
\newcommand{\NP}[1]{Nucl. \ Phys.}
\newcommand{\PL}[1]{Phys. \ Lett.}
\newcommand{\p}[1]{\partial}
\newcommand{\PRL}[1]{Phys.\ Rev.\ Lett. }
\newcommand{\MPL}[1] { Mod. Phys. Lett. }
\newcommand{\IJMP}[1] { Int. J. Mod. Phys. }
\begin{document}

\title{
On the Second Quantization of
M(atrix) Theory }
\author{
I.V.Volovich \\Steklov Mathematical Institute\\
Gubkin St.8, GSP-1, 117966, Moscow, Russia}
\date {$~$}
\maketitle
\begin {abstract}
The second quantization of M(atrix) theory
in the free (Boltzmannian) Fock space is considered.
It provides a possible framework to the recent Susskind
proposal that $U(N)$ supersymmetic Yang Mills theories for all $N$
might be embedded in a single dynamical system. The second quantization
of M(atrix) theory can also be useful for the study of the Lorentz symmetry
of the theory and for the consideration of processes with creation and
annihilation of D-branes.

\end {abstract}

Recently Susskind \cite{Sus} has suggested that there is the connection
between M-theory and $U(N)$ supersymmetric Yang-Mills theory not only
in the large $N$ limit but even at finite $N$. He also conjectured that
there is an embedding the super Yang-Mills theories for all $N$ in a single
dynamical system. In this note we attempt to pursue this idea by using
the second quantization of M(atrix) theory. The second quantization
would also be useful for the consideration of processes with creation and
annihilation of D-branes \cite{Vol}. Another problem which requires
the consideration of processes which change the size of matrices
is the problem of the Lorentz symmetry because the moment exchange in the
eleventh dimension means the exchange of zero-brane charge \cite{EMM}.

The bosonic part of
M(atrix) theory \cite{BFSS} is described by the following
Lagrangian
\begin{eqnarray}
\label{1}
L=-\frac{1}{2}{\rm tr}(\dot{ X}^{i}\dot{X}^i+\frac{1}{2}[X^i,X^j]^2)
\end{eqnarray}
where $i,j=1,...,9$. We will consider the case of the gauge group
$O(N)$. In this case  $X^i$ are real matrices satisfying $X_{ab}^i=
-X_{ba}^i$. The quantum  Hamiltonian corresponding
to the Lagrangian (\ref{1}) is
\begin{eqnarray}
\label{3}
H=-\frac{1}{2}\sum_{a<b}
\frac{\partial^2}{\partial X^{i2}_{ab}}
+\frac{1}{2}\sum_{a,b,c,d}(X_{ab}^i X_{bc}^jX_{cd}^i X_{da}^j -
X^i_{ab}X^j_{bc}X^j_{cd}X^i_{da})
\end{eqnarray}
We will quantize the theory
(\ref{3}) in the so called free (or Boltzmannian) Fock space. About this
formalism and its applications to the large $N$ limit see
\cite{Dou,GG,Sin,ALV}. The master field for QCD has been constructed
in \cite{AV1}. Let
us consider the creation and annihilation operators satisfying the following
relations
\begin{eqnarray}
\label{4}
\psi({\bf
x})\psi^*({\bf y})= \delta^{(9)}({\bf x} - {\bf y})
\end{eqnarray}
where $ {\bf x},{\bf y}\in R^9$. The second quantized Hamiltonian
corresponding to (\ref{3}) we will take in the following form
\begin{eqnarray}
\label{5}
H=\frac{1}{2} \int d{\bf u}
:\frac{\partial}{\partial
u^i}\psi^*({\bf u})\frac{\partial} {\partial u^i}\psi ({\bf u})
\frac{1}{1-K}:
\end{eqnarray}
$$
+\frac{1}{2}\int
d{\bf u}_1 d{\bf u}_2 d{\bf u}_3 d{\bf u}_4 (u_1^iu_2^ju_3^iu_4^j
$$
$$
-u_1^iu_2^ju_3^ju_4^i):
\psi^* ({\bf u}_1)\psi^* ({\bf u}_2)\psi^* ({\bf u}_3)\psi^* ({\bf u}_4)
\psi ({\bf u}_4)\psi ({\bf u}_3)\psi ({\bf u}_2)\psi ({\bf u}_1)
\frac{1}{1-K}:
$$
Here   ${\bf u},{\bf u}_1,...\in R^9$,
\begin{eqnarray}
\label{6}
K=\int
d{\bf u}_1 d{\bf u}_2 d{\bf u}_3 d{\bf u}_4
\psi^* ({\bf u}_1)\psi^* ({\bf u}_2)\psi^* ({\bf u}_3)\psi^* ({\bf u}_4)
\psi ({\bf u}_4)\psi ({\bf u}_3)\psi ({\bf u}_2)\psi ({\bf u}_1),
\end{eqnarray}
\begin{eqnarray}
\label{7}
:\frac{\partial}{\partial
u^i}\psi^*({\bf u})\frac{\partial} {\partial u^i}\psi ({\bf u})
\frac{1}{1-K}:=
\sum_{n=0}^{\infty}
\int d{\bf v}_1d{\bf v}_2...d{\bf v}_{4n}\psi^*({\bf v}_1)\psi^*({\bf v}_2)
\end{eqnarray}
$$
...\psi^*({\bf v}_{4n})
(\frac{\partial}{\partial
u^i}\psi^*({\bf u})\frac{\partial} {\partial u^i}\psi ({\bf u})
)\psi({\bf v}_{4n})...\psi({\bf v}_2)\psi({\bf v}_1)
$$
Let us consider the following  multiparticle vector
\begin{eqnarray}
\label{8}
\Phi_N=\int F_N(\{{\bf X}_{mn}\})
\prod_{a,b,c,d}(
\psi^*({\bf X}_{ab})
\psi^*({\bf X}_{bc})
\psi^*({\bf X}_{cd})
\psi^*({\bf X}_{da}))
\prod_{
 m<n}
 d{\bf X}_{mn}|0\rangle
\end{eqnarray}
Here the wave function $F_N$ is taken to be the symmetric function of the
variables $X^i_{mn}$, the product of noncommutative operators
$
\psi^*({\bf X}_{ab})...
$ is in arbitrary order and we assume $a\neq b, b\neq c, c\neq d, d\neq a$.
By acting (\ref{5}) to the vector (\ref{8}) we will get the Hamiltonian
of the form (\ref{3}):
\begin{eqnarray}
\label{9}
H=-\frac{1}{2}C_N\sum_{a<b}
\frac{\partial^2}{\partial X^{i2}_{ab}}
+\frac{1}{2}\sum_{a,b,c,d}(X_{ab}^i X_{bc}^jX_{cd}^i X_{da}^j -
X^i_{ab}X^j_{bc}X^j_{cd}X^i_{da})
\end{eqnarray}
where $C_N$ is equal to the number of $X_{ab}$ in the product
$\psi^*({\bf X}_{ab})...$ in (\ref{8}).

The Hamiltonian
(\ref{5}) doesn't depend on $N$ and it describes the $O(N)$ matrix quantum
mechanics for any $N$. We can fix $N$ if we take a vector of the form
(\ref{8}). In this sense we get the embedding of $O(N)$ theory to the single
theory for all $N$.  Applications of the Hamiltonian (\ref{5}) will be
considered in another work.

The analogous second quantization of the matrix string theory
\cite{Mot,BS,DVV} leads to the third quantization of theory of strings.

A string field theory approach to D-branes based on the connection
between Witten's string field theory and matrix models \cite{AV1}
has been considered in \cite{Are}.

I thank I.Ya.Aref'eva for useful discussions of the improved version
of this note.

$$~$$
%%%%%%%%%%%%%%%%%%%%%%%%%%%%

\end{document}